\documentclass[12pt]{article}
\pdfoutput=1
\usepackage{geometry,enumerate,amsmath,amssymb}
\usepackage{fullpage}
\usepackage{graphicx}
\usepackage{bm}
\usepackage[colorlinks=false, urlcolor=blue, linkcolor=red, hidelinks]{hyperref}
\numberwithin{equation}{section}
\newcommand{\be}{\begin{equation}}
\newcommand{\ee}{\end{equation}}
\newcommand{\bea}{\begin{eqnarray}}
\newcommand{\eea}{\end{eqnarray}}
\renewcommand{\epsilon}{\varepsilon}

\begin{document}
\title{Hyperbolic monopole data}
\author{
  Paul Sutcliffe\\[10pt]
 {\em \normalsize Department of Mathematical Sciences,}\\
 {\em \normalsize Durham University, Durham DH1 3LE, United Kingdom.}\\
 {\normalsize Email:  p.m.sutcliffe@durham.ac.uk}
}

\date{November 2025}

\maketitle
\begin{abstract}
  It is known that hyperbolic monopoles, with a particular value of the curvature, can be obtained from ADHM instanton data that satisfies additional constraints. Here this data is reformulated in terms of a triplet of real matrices that satisfy a set of quartic equations, with solutions associated with representations of $\mathfrak{su}(2)$. Many of the known examples of hyperbolic monopoles can easily be recovered in this formulation by evaluating Nahm data for Euclidean monopoles at the centre of its domain. Toda reductions of Nahm's equation correspond to cyclic Euclidean monopoles, and this is adapted to the hyperbolic setting to obtain solutions, even when the corresponding Nahm data is not tractable. A new family of charge 4 hyperbolic monopoles with square symmetry is presented as an example.
\end{abstract}

\newpage

\section{Introduction}\quad
This paper concerns $SU(2)$ hyperbolic monopoles \cite{At}, namely, solutions of the Bogomolny equation for a Yang-Mills-Higgs gauge theory in three-dimensional hyperbolic space $\mathbb{H}^3$. This equation is integrable for any value of the curvature, including the flat space Euclidean limit, and has a moduli space of monopole solutions of dimension $4N-1$, for positive monopole charge $N.$

In the zero curvature Euclidean limit, the Nahm transform \cite{Nahm} provides a correspondence between charge $N$ monopoles and Nahm data, which are a triplet of $N\times N$ matrices defined in an interval and satisfying an ordinary differential equation (Nahm's equation) together with some boundary conditions at the ends of the interval. For generic values of the curvature, a hyperbolic analogue of the Nahm transform is not known. Nevertheless, for one specific value of the curvature (determined by the asymptotic magnitude of the Higgs field), charge $N$ monopoles may be identified with charge $N$ circle-invariant Yang-Mills instantons in 4-dimensional Euclidean space \cite{At}. For this tuned value of the curvature the ADHM construction of instantons \cite{ADHM} may be adapted to the hyperbolic monopole setting, with constraints on the ADHM instanton data identified \cite{MS} to impose the circle invariance in a way that is compatible with the natural action of rotations in the ball model of $\mathbb{H}^3$. From now on, any reference to hyperbolic monopoles refers to this specific tuned value for the curvature of $\mathbb{H}^3$.

In Section 3 this constrained quaternionic ADHM data is reformulated in terms of a triplet of $N\times N$ matrices, to highlight a resemblance to Nahm data. Indeed it is shown in Section 4 that many of the known examples of hyperbolic monopole data can easily be recovered in this formulation by evaluating Nahm data at the centre of its domain. The triplet of matrices satisfy a set of quartic equations, with solutions associated with representations of $\mathfrak{su}(2)$. The relevant representations are identified for a number of illustrative examples. In Section 5 the similarity with Nahm data is exploited by adapting the known Toda reduction of Nahm's equation, to impose cyclic symmetry, to the hyperbolic setting. A new family of charge 4 hyperbolic monopoles with square symmetry is presented as an example of this approach.

\section{Hyperbolic monopoles, JNR and ADHM}\quad
This section contains a brief review of some material on hyperbolic monopoles that will be required in the subsequent sections.

The Bogomolny equation for hyperbolic monopoles is
\be
D\Phi=*F,
\label{Bog}
\ee
where $F$ is the field strength of an $SU(2)$ gauge
potential, and $D\Phi$ is the covariant derivative of an adjoint Higgs
field $\Phi$. The Hodge dual $*,$ 
is defined on $\mathbb{H}^3$ with curvature
$-1,$ which is the particular tuned value mentioned in the introduction, given that the
magnitude of the Higgs field, $|\Phi|=\sqrt{-\frac{1}{2}\mbox{Tr}\,\Phi^2},$ is fixed to have the asymptotic value
$\frac{1}{2}$ at the boundary $\partial \mathbb{H}^3.$
The unit ball model of $\mathbb{H}^3$ will be used, with coordinates $X=(X_1,X_2,X_3)$, where $|X|^2=X_1^2+X_2^2+X_3^2<1.$ The monopole charge $N$ is the degree of the map between unit two-spheres $2\Phi\vert_{|X|=1}:S^2\mapsto S^2.$ For later use, it is noted that the energy density of the hyperbolic monopole is equal to the Laplace-Beltrami operator acting on $|\Phi|^2$.

In Sections 4 and 5 it will be helpful to describe a hyperbolic monopole by its spectral curve \cite{At}, so the salient features of this approach will now be reviewed. This is a holomorphic description of the monopole that encodes a collection of geodesics in $\mathbb{H}^3$, with a particular property, using the twistor correspondence. The mini-twistor space of a real 3-manifold with constant curvature is the complex surface given by the space of its oriented geodesics.
An oriented geodesic in $\mathbb{H}^3$ may be specified by giving two distinct points on the boundary two-sphere $|X|=1$, say $\hat X_{\rm start}$ and $\hat X_{\rm end}$, to label the geodesic that starts at $\hat X_{\rm start}$ and ends at $\hat X_{\rm end}$. Let $\zeta$ be the Riemann sphere coordinate corresponding to the point $\hat X_{\rm end}$ at the end of the geodesic and let $\eta$ be the Riemann sphere coordinate corresponding to the point that is antipodal to the starting point $\hat X_{\rm start}$ of the geodesic. In other words, the pair of $\mathbb{CP}^1$ coordinates $(\eta,\zeta)$ specify the oriented geodesic that starts at $-1/\bar\eta$ and ends at $\zeta$. The fact that the start and end points of the geodesic cannot coincide requires that the anti-diagonal, where $\zeta\bar\eta=-1$, must be excluded. Hence the mini-twistor space of $\mathbb{H}^3$ is $\mathbb{CP}^1 \times\mathbb{CP}^1,$ with the anti-diagonal removed.

The particular property that is required of a geodesic is that Hitchin's equation \cite{Hi1}
\be
(D_u-i\Phi)Y=0,
\ee
along the geodesic has a solution that decays at both ends, where $Y$ is a complex two-component scalar and $u$ is arc length along the geodesic. Such a geodesic is called spectral, and the spectral curve is the collection of all the spectral geodesics, using their mini-twistor space description. The upshot is that the spectral curve is an algebraic curve in mini-twistor space with bidegree $(N,N)$
that takes the form
\be \sum_{i=0,j=0}^N c_{ij}\eta^i\zeta^j=0. 
\label{gensc}
\ee
The complex constants $c_{ij}$ satisfy the reality condition
$\bar c_{ij}=(-1)^{N+i+j}c_{N-j,N-i}$,
which follows from reversing the orientation of the geodesic.
Given an algebraic curve of the form (\ref{gensc}), satisfying the reality condition, it is generally a difficult task to determine if this is a spectral curve, as the conditions that it must satisfy involve imposing constraints on the constants $c_{ij}$ that involve relations between integrals of holomorphic differentials around particular cycles of the curve.
However, for a single monopole with position $X$ in the unit ball the spectral curve is simply
\be
2\eta\zeta(X_1-iX_2)+\zeta(1+|{ X}|^2-2X_3)-\eta(1+|{ X}|^2+2X_3)-2(X_1+iX_2)=0,
\label{star}
\ee
capturing all the geodesics that pass through the point $X.$
Roughly speaking, for charge $N$ the physical interpretation is to think of the spectral curve as corresponding to all the geodesics that pass through any of the positions of the $N$ monopoles.

The simplest method to obtain some hyperbolic monopoles as circle-invariant instantons is to use the subclass of JNR instantons \cite{JNR} given by a harmonic ansatz \cite{CF}. Invariance under the circle action is obtained by placing the $N+1$ poles of the harmonic ansatz on the two-sphere that is the fixed point set of the action. The result is an explicit formula for the hyperbolic monopole fields \cite{MS} in terms of the free data of $N+1$ distinct points on the sphere, each with an associated real positive weight. For $N\in\{1,2,3\}$ this construction of JNR monopoles yields the full $(4N-1)$-dimensional moduli space of hyperbolic monopoles, but for $N>3$ it provides only a $(3N+2)$-dimensional subspace.

 Let the free data for a JNR monopole be $\lambda_j,\gamma_j$, for $j=0,1,...,N$, where $\lambda_j$ are any set of real positive constants (the weights), and $\gamma_j$ are the Riemann sphere coordinates of any set of $N+1$ distinct points on the sphere (the poles). There is a compact formula for the spectral curve of a JNR monopole \cite{BCS}, given by
\be
\sum_{j=0}^N\lambda_j^2\mathop{\prod_{k=0}^N}_{k\ne j}
(\zeta-\gamma_k)(1+\eta\bar\gamma_k)=0.
\label{JNRspectralcurve}
\ee
If the weights are chosen to be $\lambda_j^2=1+|\gamma_j|^2$, then they are termed canonical, and in this case the hyperbolic monopole inherits the rotational symmetry of the $N+1$ points on the sphere.

To obtain hyperbolic monopoles beyond the JNR class requires access to the full moduli space of $SU(2)$ charge $N$ Yang-Mills instanton in $\mathbb{R}^4$, so that circle invariance may then be imposed. This is possible via the ADHM construction of instantons \cite{ADHM}, which is a correspondence between instantons and a matrix of quaternions, known as ADHM data. For a charge $N$ instanton this data consists of a symmetric $N\times N$ matrix of quaternions $M$, together with a non-zero $N$-component row vector of quaternions, $L$. These are arranged into the ADHM matrix \be
\widehat M = \begin{pmatrix} L \\ M 
\end{pmatrix},
\ee
that is required to satisfy the condition that
$\widehat M^\dagger \, \widehat M$ is a real invertible $N\times N$ matrix, where
$^\dagger$ denotes quaternionic conjugate transpose.
Circle invariance of the ADHM data is imposed by requiring that the following three additional constraints \cite{MS} are satisfied,
\bea
&& M \mbox{\ is pure quaternion,}\label{con1}\\
&& \widehat M^\dagger \, \widehat M \mbox{\ is the identity matrix,}\label{con2}\\
&& LM=\mu L \,, \mbox{\ where $\mu$ is a pure quaternion.}\label{con3} 
\eea
The quantity $|\mu|\in[0,1)$ has a physical interpretation as the ratio of the magnitude of the Higgs field $|\Phi|$ evaluated at the origin $X=0$ and at the boundary  $|X|=1$ of $\mathbb{H}^3$, where it takes the value $\frac{1}{2}.$

From the constrained ADHM data, the Higgs field, viewed as a pure quaternion, is given by
\be
\Phi=\frac{1}{2}\Psi^\dagger \begin{pmatrix}-\mu & L\\ -L^\dagger &  M\end{pmatrix}
  \Psi,\label{Higgs}
  \ee
  where $\Psi$ is a unit length $(N+1)$-component column vector of quaternions that solves the linear system
  \be
  (L^\dagger,X-M)\Psi=0,
  \ee
  with the spatial coordinate in the unit ball written as a pure quaternion
$X={ i}X_1+{ j}X_2+{ k}X_3$.  
Regularity requires that the real matrix $L^\dagger L-(M-X)^2$ is non-singular.

There is a simple formula \cite{Su1} for the spectral curve in terms of the constrained ADHM data, given by
\be
\mbox{det}\big(\eta\zeta(M_1-iM_2)+\zeta(1-M_3)-\eta(1+M_3)-(M_1+iM_2)\big)=0,
\label{sc}
\ee
where the triplet $M_1,M_2,M_3$ of real symmetric matrices are extracted from $M$ by writing 
$M={ i}M_1+{ j}M_2+{ k}M_3.$ Note that (\ref{sc}) does not involve $L$, which suggests that there should be a reformulation of the constraints (\ref{con2}) and (\ref{con3}) that avoids the introduction of $L$ and defines hyperbolic monopole data solely in terms of the pure quaternion symmetric matrix $M$, or equivalently the triplet of real symmetric matrices $M_1,M_2,M_3$. This is indeed the case and is presented in the next section.

\section{A reformulation of hyperbolic monopole data}\quad
To be hyperbolic monopole data, the pure quaternion symmetric matrix $M$ must satisfy the equation
\be
(M^2+1)(M^2+\alpha^2)=0,
\label{quartic}
\ee
for some $\alpha^2\in[0,1).$  This may be rewritten as 
  \be
  \bigg(\frac{1+M^2}{1-\alpha^2}\bigg)^2=
    \frac{1+M^2}{1-\alpha^2},
  \ee
so the condition is that $P=(1+M^2)/(1-\alpha^2)$ is a hermitian projector, and in fact it is required to have rank one.

Recall that a quaternionic matrix $M$ has a right eigenvector $v$ with eigenvalue $\lambda$ if
  \be Mv=v\lambda.\label{ev}\ee
An eigenvalue is called standard if it has
the form $\lambda=\lambda_0+{ i}\lambda_1$, where $\lambda_0$ and $\lambda_1$ are real with $\lambda_1\ge 0$. Any eigenvalue belongs to the equivalence class of a standard eigenvalue, with the equivalence relation being conjugation by a unit quaternion. An $N\times N$ quaternionic matrix has $N$ standard eigenvalues, counted with multiplicity \cite{Lee}.     

  Applying (\ref{quartic}) to $v$ and using (\ref{ev}) gives
  \be
  (\lambda^2+1)(\lambda^2+\alpha^2)=0.
  \ee
  Let $v$ be a unit eigenvector corresponding to the standard eigenvalue
  $\lambda={ i}\alpha.$ 
As $Pv=v$ then $P$ projects onto the space spanned by $v$, hence $P=vv^\dagger.$
  Defining
  \be
  L=v^\dagger\sqrt{1-\alpha^2},\label{defL}
  \ee
  then $1+M^2=(1-\alpha^2)P=(1-\alpha^2)vv^\dagger=L^\dagger L$.
  This shows that (\ref{con2}) is satisfied because
  $\widehat M^\dagger \, \widehat M = L^\dagger L-M^2=1$.
  Taking the quaternionic conjugate transpose of (\ref{ev}) with $\lambda={ i}\alpha$ and using (\ref{defL}) gives $LM={ i}\alpha L$, therefore (\ref{con3}) is satisfied with $\mu={ i}\alpha.$ This confirms that hyperbolic monopole data yields ADHM data satisfying the required constraints.

In terms of the triplet of real symmetric matrices $M_1,M_2,M_3$,
the calculation of the quaternionic eigenvector (\ref{ev}) with complex eigenvalue $\lambda=i\alpha$ can be reduced to complex linear algebra by splitting the $2N$-component complex eigenvector $w$
\be
\begin{pmatrix} iM_1 & M_2+iM_3 \\ -M_2+iM_3 & -iM_1 \end{pmatrix}
w=i\alpha w,
\label{complexeig}
\ee
into a pair of $N$-component complex vectors $w^t=(w_+^t,w_-^t)$, and then
$v=w_+-\bar w_-j$ is the required quaternionic eigenvector \cite{Lee}.

To write (\ref{quartic}) in terms of the real matrices $M_i$, for $i=1,2,3$, define the following quadratic combinations,
\be S=M_iM_i, \quad\quad  A_i=\varepsilon_{ijk}M_jM_k,
\label{defSA}
\ee
where $\varepsilon_{ijk}$ is the totally antisymmetric tensor and repeated indices are summed over the values $1,2,3$.
Note that $S$ is a symmetric matrix, whereas $A_i$ are antisymmetric. A basis can always be chosen so that $S$ is diagonal, and it will be convenient to work in such a basis from now on. 
With this notation (\ref{quartic}) becomes the set of equations
\be
(S-1)(S-\alpha^2)=A_iA_i,\quad\quad
(1+\alpha^2)A_i+\varepsilon_{ijk}A_jA_k=SA_i+A_iS.
   \label{quadi}
   \ee
   Real representations of $\mathfrak{su}(2)$ provide solutions of these equations, as follows. Let $r_1,r_2,r_3$ be a real basis for the $d$-dimensional irreducible representation $\underline{d}$ of $\mathfrak{su}(2)$, normalized so that
$\varepsilon_{ijk}r_jr_k=2r_i.$ For the moment take $d=N$, with $N$ odd, so that a real irreducible representation exists. Recall that the quadratic Casimir for $\underline{d}$ is proportional to the identity matrix and is given by $r_1^2+r_2^2+r_3^2=1-d^2.$ 
It is easy to verify that a solution to (\ref{quadi}) is obtained by setting
\be
A_i=\frac{(1-\alpha^2)}{2d}r_i,\label{ftor}
\ee
providing $S$ is proportional to the identity matrix with the scale factor
       \be
       S=\frac{1}{2d}\left(d+1+\alpha^2(d-1)\right).
       \label{f4}
       \ee
       The hyperbolic monopole data $M_i$ may therefore be thought of as a kind of prebasis for the representation, in the sense that after a suitable normalization $r_1$ is given by $[M_2,M_3]$, etc.       

       This form of solution clearly generalizes to the case of a reducible $N$-dimensional representation, where the above formulae apply to each irreducible component $\underline{d}$, but the value of $\alpha$ must be the same for all the irreducible pieces. In this case $S$ might not be proportional to the identity matrix, but rather will have a block structure with each block being a constant times the $d\times d$ identity matrix, mirroring the structure of the quadratic Casimir.
 Real irreducible representations of $\mathfrak{su}(2)$ require that the dimension is odd or a multiple of 4. With a slight abuse of notation, the even irreducible representations over the complex numbers will be used, so that a real irreducible representation of dimension $2d$, with $d$ even, will be denoted by $\underline{d}\oplus\underline{d}.$ The advantage of this is that the formula for the quadratic Casimir, together with (\ref{ftor}) and (\ref{f4}) continue to apply, although the block structure will be as a $2d\times2d$ block, rather than a $d\times d$ block. Some illustrative examples will be provided in the following section, and the associated representations of $\mathfrak{su}(2)$ identified. 
       
\section{Recycling Nahm data}\quad
The purpose of this section is twofold. Firstly, it will be shown how several known examples of hyperbolic monopoles fit into the formalism introduced in Section 3, with the representations of $\mathfrak{su}(2)$ determined in each case. Secondly, it is found that the hyperbolic monopole data for these examples is easily recovered from known Nahm data through a simple process that will be referred to as recycling Nahm data.

   Nahm data \cite{Nahm} for charge $N$ monopoles in Euclidean space consists of three $N\times N$ antihermitian matrices $T_i(s),$ that depend on a real variable $s.$ The matrices are smooth for $s\in(-1,1)$ and satisfy Nahm's equation,
   \be
   \frac{dT_i}{ds}=\varepsilon_{ijk}T_jT_k.
   \ee
   At the ends of the interval, $s=\pm 1$, the matrices have simple poles, with the matrix residues forming the irreducible $N$-dimensional representation of $\mathfrak{su}(2)$. There is also a reflection property about the centre of the interval, $T_i(-s)=T_i^t(s)$, although often a basis is chosen in which this reflection property is not manifest.
   
 The recycling of Nahm data is based on the observation that many examples of hyperbolic monopole data $M_i$ that satisfy (\ref{quadi}) can easily be obtained from known Nahm data. The recycling process is simply to evaluate the Nahm data at the centre of the interval and scale by an appropriate pure imaginary constant. Explicitly, set $M_i=-i\beta T_i(0)$, where $\beta$ is a real positive constant. Note that the reflection property ensures that $T_i(0)$ is a symmetric matrix, and as $T_i(0)$ is antihermitian then $-iT_i(0)$ is a real symmetric matrix.
The following examples illustrate the details of the recycling process.

   The Nahm data of a charge 3 monopole with tetrahedral symmetry is given by \cite{HMM}
   \be
   T_1=\begin{pmatrix}
   0&0&0\\0&0&-x+iy\\0&x+iy&0
   \end{pmatrix},
   T_2=\begin{pmatrix}
   0&0&x+iy\\0&0&0\\-x+iy&0&0
   \end{pmatrix},
   T_3=\begin{pmatrix}
   0&-x+iy&0\\x+iy&0&0\\0&0&0
   \end{pmatrix},
   \ee
   where $x$ and $y$ are real functions of $s$ that are given in terms of an elliptic function $\wp$ and its derivative. The properties of $x$ and $y$ that are relevant here is that is $x$ is an odd function of $s$ and $y$ is an even function. Evaluating at $s=0$, and absorbing the value of $y(0)$ into a redefinition of $\beta$, yields
 \be
   M_1=\beta\begin{pmatrix}
   0&0&0\\0&0&1\\0&1&0
   \end{pmatrix},\
   M_2=\beta\begin{pmatrix}
   0&0&1\\0&0&0\\1&0&0
   \end{pmatrix},\
   M_3=\beta\begin{pmatrix}
   0&1&0\\1&0&0\\0&0&0
   \end{pmatrix}.
   \label{tet3}
   \ee  
   The required value of $\beta$ is found by taking the definitions (\ref{defSA}) and using (\ref{tet3}) to calculate that
   \be
   A_1=\beta^2
   \left(\begin{array}{ccc}
0 & 0 & 0 
\\
 0 & 0 & -1 
\\
 0 & 1 & 0 
   \end{array}\right),\
   A_2=\beta^2
   \left(\begin{array}{ccc}
0 & 0 & 1 
\\
 0 & 0 & 0 
\\
 -1 & 0 & 0 
   \end{array}\right),\
   A_3=\beta^2
   \left(\begin{array}{ccc}
0 & -1 & 0 
\\
 1 & 0 & 0 
\\
 0 & 0 & 0 
\end{array}\right).
\ee
The eigenvalues of $iA_3$ are $\pm \beta^2,0,$ and 
$r_i=2A_i/\beta^2$ form the representation $\underline{3}.$
Comparing this normalization of $A_i$ with (\ref{ftor}) gives
$\beta^2=(1-\alpha^2)/3$.
The final step is to confirm that $S$ is proportional to the identity matrix, and indeed $S=2\beta^2=\frac{2}{3}(1-\alpha^2).$ Requiring this to be the correct scale factor (\ref{f4}) determines that $\alpha=0$ and hence $\beta=\frac{1}{\sqrt{3}}.$ This reproduces the constrained ADHM data from \cite{MS} for the tetrahedrally symmetric charge 3 hyperbolic monopole, with $L$ recovered using (\ref{defL}) together with (\ref{complexeig}).

An example with a reducible representation is obtained by taking the Nahm data of a charge 5 monopole with octahedral symmetry \cite{HS2}. Evaluating the Nahm data at $s=0$, and again absorbing an appropriate constant within a redefinition of $\beta$ for convenience (as will be done from now on without comment), gives
\bea
& &M_1=\beta
\left(\begin{array}{ccccc}
0 & 0 & 0 & 0 & 0 
\\
 0 & 0 & 0 & 1 & \sqrt{3} 
\\
 0 & 0 & 0 & 0 & 0 
\\
 0 & 1 & 0 & 0 & 0 
\\
 0 & \sqrt{3} & 0 & 0 & 0 
  \end{array}\right),\quad
M_2=\beta
\left(\begin{array}{ccccc}
0 & 0 & 0 & 0 & 0 
\\
 0 & 0 & 0 & 0 & 0 
\\
 0 & 0 & 0 & -1 & \sqrt{3} 
\\
 0 & 0 & -1 & 0 & 0 
\\
 0 & 0 & \sqrt{3} & 0 & 0 
  \end{array}\right),\nonumber\\
& &M_3=\beta
\left(\begin{array}{ccccc}
0 & 0 & 0 & -2 & 0 
\\
 0 & 0 & 0 & 0 & 0 
\\
 0 & 0 & 0 & 0 & 0 
\\
 -2 & 0 & 0 & 0 & 0 
\\
 0 & 0 & 0 & 0 & 0 
\end{array}\right).
\label{oct5}
\eea
To verify that an appropriate value of $\beta$ gives a solution to (\ref{quadi}), the result of the calculation of the matrices (\ref{defSA}) is
\bea
& &A_1=\beta^2
\left(\begin{array}{ccccc}
0 & 0 & -2 & 0 & 0 
\\
 0 & 0 & 0 & 0 & 0 
\\
 2 & 0 & 0 & 0 & 0 
\\
 0 & 0 & 0 & 0 & 0 
\\
 0 & 0 & 0 & 0 & 0 
\end{array}\right),\quad
A_2=\beta^2
\left(\begin{array}{ccccc}
0 & -2 & 0 & 0 & 0 
\\
 2 & 0 & 0 & 0 & 0 
\\
 0 & 0 & 0 & 0 & 0 
\\
 0 & 0 & 0 & 0 & 0 
\\
 0 & 0 & 0 & 0 & 0 
\end{array}\right),\nonumber\\
& &A_3=\beta^2
\left(\begin{array}{ccccc}
0 & 0 & 0 & 0 & 0 
\\
 0 & 0 & 2 & 0 & 0 
\\
 0 & -2 & 0 & 0 & 0 
\\
 0 & 0 & 0 & 0 & 0 
\\
 0 & 0 & 0 & 0 & 0 
\end{array}\right),\quad
S=\beta^2
\left(\begin{array}{ccccc}
4 & 0 & 0 & 0 & 0 
\\
 0 & 4 & 0 & 0 & 0 
\\
 0 & 0 & 4 & 0 & 0 
\\
 0 & 0 & 0 & 6 & 0 
\\
 0 & 0 & 0 & 0 & 6 
\end{array}\right).
\eea
The eigenvalues of $iA_3$ are $\pm 2\beta^2$ and $0$ (with multiplicity 3), with $r_i=A_i/\beta^2$ a basis for the representation
$\underline{3}\oplus\underline{1}\oplus\underline{1}$. Comparing this normalization with (\ref{ftor}) requires that $6\beta^2=1-\alpha^2$, and equating $S$ to the expression (\ref{f4}) for the $d=3$ block and the two $d=1$ blocks requires that $2+\alpha^2=12\beta^2$ and $6\beta^2=1$, hence $\beta=\frac{1}{\sqrt{6}}$ and $\alpha=0.$

Calculating the spectral curve of this data using (\ref{sc}) gives
\be
(\eta-\zeta)\left((\eta^4-1)(\zeta^4-1)+8\eta\zeta(\eta^2+\zeta^2)\right)=0.
\ee
This hyperbolic monopole is already known within the JNR formulation, being
obtained from the JNR data of 6 points (with canonical weights) on the unit sphere at the vertices of an octahedron. Applying (\ref{JNRspectralcurve}) to this JNR data confirms that the same spectral curve is indeed reproduced.

Note that the residues at the poles of the Nahm data that has been recycled form the representation $\underline{5}$, but this does not emerge as the representation associated with the hyperbolic data, which is
$\underline{3}\oplus\underline{1}\oplus\underline{1}$.

Next, consider an example in which a representation that is a multiple of 4 appears, written as $\underline{d}\oplus\underline{d},$ for $d$ even. Evaluating the Nahm data for a charge 7 monopole with dodecahedral symmetry \cite{HS2} at $s=0$ produces 
\bea
& &M_1=
\beta \left(\begin{array}{ccccccc}
0 & 0 & 0 & 0 & 0 & 0 & -1 
\\
 0 & 0 & 0 & 0 & 0 & 0 & 1 
\\
 0 & 0 & 0 & 0 & 1 & 2 & 0 
\\
 0 & 0 & 0 & 0 & 1 & 0 & 0 
\\
 0 & 0 & 1 & 1 & 0 & 0 & 0 
\\
 0 & 0 & 2 & 0 & 0 & 0 & 0 
\\
 -1 & 1 & 0 & 0 & 0 & 0 & 0 
\end{array}\right),\quad
M_2=\beta \left(\begin{array}{ccccccc}
0 & 0 & 0 & 0 & -1 & 0 & 0 
\\
 0 & 0 & 0 & 0 & -1 & 2 & 0 
\\
 0 & 0 & 0 & 0 & 0 & 0 & 1 
\\
 0 & 0 & 0 & 0 & 0 & 0 & -1 
\\
 -1 & -1 & 0 & 0 & 0 & 0 & 0 
\\
 0 & 2 & 0 & 0 & 0 & 0 & 0 
\\
 0 & 0 & 1 & -1 & 0 & 0 & 0 
\end{array}\right),\nonumber\\
& &
M_3=
\beta \left(\begin{array}{ccccccc}
0 & 0 & 0 & 0 & 2 & 0 & 0 
\\
 0 & 0 & 0 & 0 & 0 & 0 & 0 
\\
 0 & 0 & 0 & 0 & 0 & 0 & 0 
\\
 0 & 0 & 0 & 0 & 0 & 0 & -2 
\\
 2 & 0 & 0 & 0 & 0 & 0 & 0 
\\
 0 & 0 & 0 & 0 & 0 & 0 & 0 
\\
 0 & 0 & 0 & -2 & 0 & 0 & 0 
\end{array}\right).
\eea
The associated matrices are 
\bea
& &A_1=\beta^2
\left(\begin{array}{ccccccc}
0 & 2 & 0 & 0 & 0 & 0 & 0 
\\
 -2 & 0 & 0 & 0 & 0 & 0 & 0 
\\
 0 & 0 & 0 & -2 & 0 & 0 & 0 
\\
 0 & 0 & 2 & 0 & 0 & 0 & 0 
\\
 0 & 0 & 0 & 0 & 0 & 0 & 0 
\\
 0 & 0 & 0 & 0 & 0 & 0 & 0 
\\
 0 & 0 & 0 & 0 & 0 & 0 & 0 
\end{array}\right),\quad
A_2=\beta^2
\left(\begin{array}{ccccccc}
0 & 0 & 2 & 0 & 0 & 0 & 0 
\\
 0 & 0 & 0 & 2 & 0 & 0 & 0 
\\
 -2 & 0 & 0 & 0 & 0 & 0 & 0 
\\
 0 & -2 & 0 & 0 & 0 & 0 & 0 
\\
 0 & 0 & 0 & 0 & 0 & 0 & 0 
\\
 0 & 0 & 0 & 0 & 0 & 0 & 0 
\\
 0 & 0 & 0 & 0 & 0 & 0 & 0 
\end{array}\right),\nonumber\\
& &
A_3=\beta^2
\left(\begin{array}{ccccccc}
0 & 0 & 0 & 2 & 0 & 0 & 0 
\\
 0 & 0 & -2 & 0 & 0 & 0 & 0 
\\
 0 & 2 & 0 & 0 & 0 & 0 & 0 
\\
 -2 & 0 & 0 & 0 & 0 & 0 & 0 
\\
 0 & 0 & 0 & 0 & 0 & 0 & 0 
\\
 0 & 0 & 0 & 0 & 0 & 0 & 0 
\\
 0 & 0 & 0 & 0 & 0 & 0 & 0 
\end{array}\right),\quad
S=\beta^2
\left(\begin{array}{ccccccc}
6 & 0 & 0 & 0 & 0 & 0 & 0 
\\
 0 & 6 & 0 & 0 & 0 & 0 & 0 
\\
 0 & 0 & 6 & 0 & 0 & 0 & 0 
\\
 0 & 0 & 0 & 6 & 0 & 0 & 0 
\\
 0 & 0 & 0 & 0 & 8 & 0 & 0 
\\
 0 & 0 & 0 & 0 & 0 & 8 & 0 
\\
 0 & 0 & 0 & 0 & 0 & 0 & 8 
\end{array}\right).
\eea
The eigenvalues of $iA_3$ are $\pm 2\beta^2$ (each with multiplicity 2) and $0$ (with multiplicity 3), with $r_i=A_i/(2\beta^2)$ a basis for the representation
$\underline{2}\oplus\underline{2}\oplus\underline{1}\oplus\underline{1}\oplus\underline{1}$. Comparing this normalization with (\ref{ftor}) requires that $8\beta^2=1-\alpha^2$, and equating $S$ to the expression (\ref{f4}) for the two $d=2$ blocks and the three $d=1$ blocks requires that $3+\alpha^2=24\beta^2$ and $8\beta^2=1$, hence $\beta=\frac{1}{\sqrt{8}}$ and $\alpha=0.$
This data reproduces the constrained ADHM data presented in \cite{MS}, after an appropriate rotation and change of basis, as can be verified by calculating the spectral curve using (\ref{sc}). 

As an example of a one-parameter family of Nahm data, consider the tetrahedral charge 4 data \cite{HS1} that includes the charge 4 cubic monopole \cite{HMM} as a special case within the family. Evaluating this Nahm data at $s=0$ yields,
\bea
&&M_1=\beta
\left(\begin{array}{cccc}
-a  & -\sqrt{3} & 0 & -1 
\\
 -\sqrt{3} & a  & 1 & 0 
\\
 0 & 1 & -a  & -\sqrt{3} 
\\
 -1 & 0 & -\sqrt{3} & a  
\end{array}\right),\quad
M_2=\beta
\left(\begin{array}{cccc}
\sqrt{3} & a  & 1 & 0 
\\
 a  & -\sqrt{3} & 0 & 1 
\\
 1 & 0 & -\sqrt{3} & -a  
\\
 0 & 1 & -a  & \sqrt{3} 
\end{array}\right),\nonumber\\
&&M_3=\beta
\left(\begin{array}{cccc}
-2 & 0 & 0 & -a  
\\
 0 & -2 & -a  & 0 
\\
 0 & -a  & 2 & 0 
\\
 -a  & 0 & 0 & 2 
  \end{array}\right),
\eea
where the real parameter $a$ parameterizes the family, with $a=0$ the cubic case. Calculating the associated matrices gives
\be
A_1=\delta
\left(\begin{array}{cccc}
0 & 0 & 1 & 0 
\\
 0 & 0 & 0 & 1 
\\
 -1 & 0 & 0 & 0 
\\
 0 & -1 & 0 & 0 
\end{array}\right),\quad
A_2=\delta
\left(\begin{array}{cccc}
0 & 0 & 0 & 1 
\\
 0 & 0 & -1 & 0 
\\
 0 & 1 & 0 & 0 
\\
 -1 & 0 & 0 & 0 
\end{array}\right),\quad
A_3=\delta
\left(\begin{array}{cccc}
0 & 1 & 0 & 0 
\\
 -1 & 0 & 0 & 0 
\\
 0 & 0 & 0 & -1 
\\
 0 & 0 & 1 & 0 
\end{array}\right),
\ee
where $\delta=2\beta^2(2-a^2).$ Also, $S$ is proportional to the identity matrix, $S=3\beta^2(4+a^2).$

The eigenvalues of $iA_3$ are $\pm \delta$, each with multiplicity 2, with the matrices $r_i=A_i/\delta$ a basis for the representation $\underline{2}\oplus\underline{2}$, for $a\in(-\sqrt{2},\sqrt{2}),$ so that $\delta>0$. Comparing this normalization with (\ref{ftor}) requires that $8\beta^2(2-a^2)=1-\alpha^2$, and equating $S$ to the expression (\ref{f4}) with $d=2$ requires that $3+\alpha^2=12\beta^2(4+a^2),$  hence
\be
\beta=\frac{1}{\sqrt{16+a^2}}, \quad  \alpha^2=\frac{9a^2}{{16+a^2}}.
\ee
Calculating the spectral curve confirms that this data is equivalent to the tetrahedral family presented in \cite{MS}, with $a=0$ the data of the cubic charge 4 hyperbolic monopole. As $a\to \pm \sqrt{2}$ the four hyperbolic  monopoles lie on the vertices of a tetrahedron at the boundary $\partial \mathbb{H}^3$ given by $|X|=1.$

In all the examples considered so far, the matrices $A_i$ have been obtained from the representation basis $r_i$ by applying the same scaling to all three matrices. However, this need not always be the case, as the following example demonstrates.

Recycling the Nahm data of a one-parameter family of charge 3 monopoles with dihedral $D_2$ symmetry \cite{HS3} produces the following family of hyperbolic monopole data that solve (\ref{quadi}) with $\alpha=0$ for all $a\in(-1,1)$,
\be
M_1=\sqrt{\frac{1-a^2}{2}}
\left(\begin{array}{ccc}
0 & 0 & 0 
\\
 0 & 0 & 1 
\\
 0 & 1 & 0 
\end{array}\right),\quad
M_2=\sqrt{\frac{1-a^2}{2}}
\left(\begin{array}{ccc}
0 & 0 & 1 
\\
 0 & 0 & 0 
\\
 1 & 0 & 0 
\end{array}\right),\quad
M_3=a
\left(\begin{array}{ccc}
0 & 1 & 0 
\\
 1 & 0 & 0 
\\
 0 & 0 & 0 
\end{array}\right).
\ee
The spectral curve is
\be
(\eta-\zeta)(\eta^2+\zeta^2-\frac{4a^2}{1-a^2}\eta\zeta)
-ia(\eta+\zeta)(\eta^2\zeta^2-1)=0,
\ee
which confirms that this data is equivalent to the family of JNR monopoles obtained by taking 4 points (with canonical weights) on the sphere with $D_2$ symmetry \cite{BCS}. There is axial symmetry when $a=0$, and tetrahedral symmetry when $a=\pm \frac{1}{\sqrt{3}}$. In the limit as $a\to \pm 1$ there is a hyperbolic monopole at the origin and two hyperbolic monopoles at the boundary $\partial \mathbb{H}^3$, with positions $X=(0,0,\pm 1)$. This solution is associated with the basis $r_i$ for the representation $\underline{3}$, but via the anisotropic scaling
\be
A_1=\frac{a}{2}\sqrt{\frac{1-a^2}{2}}r_1,\quad
A_2=\frac{a}{2}\sqrt{\frac{1-a^2}{2}}r_2,\quad
A_3=\frac{1-a^2}{4}r_3.
\label{aniso3}
\ee
The one-parameter family of Nahm data recycled for this example sits within a more general two-parameter family of explicit Nahm data \cite{BDH2} for charge 3 monopoles with $D_2$ symmetry. This two-parameter family can also be recycled to produce a two-parameter family of hyperbolic data, with an anisotropic scaling that extends (\ref{aniso3}) so that no pair of the factors multiplying the $r_i$ need to be equal to each other. This family includes the axial charge 3 monopole at the origin, with three possible orientations in which the symmetry axis is aligned with one of the Cartesian axes. It also includes configurations in which there is a single monopole at the origin, with two monopoles at the boundary $|X|=1$ with positions $X=(\pm 1,0,0),$ or $X=(0,\pm 1,0),$ or $X=(0,0,\pm 1)$.

In Section 3 the representation $\underline{2}$ only appeared as a component of the real 4-dimensional representation $\underline{2}\oplus\underline{2}$. However, there is a degenerate situation in which the representation $\underline{2}$ can appear on its own, despite the fact that it is not real. The degenerate situation is if one of the $M_i$ vanish, say $M_3=0.$ In this case $A_1=A_2=0$, and only the basis matrix $r_3$ is required, which can be chosen to be real. Explicitly, take the basis for $\underline{2}$ to be
\be
r_1=\left(\begin{array}{cc}
i & 0\\
 0 & -i 
\end{array}\right),\quad
r_2=\left(\begin{array}{cc}
0 & i\\
 i & 0 
\end{array}\right),\quad
r_3=\left(\begin{array}{cc}
0 & -1\\
 1 & 0 
\end{array}\right),\label{basis2}
\ee
then $A_3=\frac{1}{2}(1-\alpha^2)r_3$ with $S=\frac{1}{2}(1+\alpha^2)$
solves (\ref{quadi}). This is exactly how the data for the hyperbolic charge 2 monopole, with arbitrary separation between the two monopoles, appears from recycled Nahm data. Recycling the charge 2 Nahm data \cite{BPP} yields the hyperbolic data  $M_1=-\frac{i}{2}(1-a)r_1, \ M_2=\frac{i}{2}(1+a)r_2, \ M_3=0,$ that solves (\ref{quadi}) with $\alpha^2=a^2$ for all $a\in(-1,1).$ The parameter $a$ is related to the separation between the two monopoles, with $a=0$ the axial charge two monopole. The two monopoles approach the boundary $\partial\mathbb{H}^3$ as $a\to \pm 1$. As in the previous example, this may be viewed as obtaining $A_i$ by applying anisotropic scaling factors to $r_i$, but in this degenerate situation two of the scaling factors are identically zero to produce $A_1=A_2=0.$

\section{Toda reductions}\quad
Utilising the description of the moduli space of charge $N$ Euclidean monopoles as the space of degree $N$ based rational maps \cite{Do}, and imposing cyclic $C_N$ symmetry, yields a family of geodesics \cite{HMM} in the moduli space that will be denoted $G_N^\ell,$ where $\ell$ is an integer satisfying $0\le \ell\le  N/2,$ and $N>2$, to simplify the discussion.
Within the moduli space approximation to monopole dynamics \cite{Ma1}, each of 
these geodesics corresponds to a monopole scattering process in which $N$ single monopoles approach from infinity on the vertices of a contracting regular $N$-gon. $G_N^0$ is a planar scattering in which the cyclic symmetry $C_N$ is extended to dihedral symmetry $D_N$, with the monopoles forming the axially symmetric charge $N$ monopole before emerging on the vertices of an expanding regular $N$-gon, which is rotated through an angle $\pi/N$ relative to the incoming $N$-gon. For $\ell>0$ the scattering $G_N^\ell$ results in a charge $\ell$ axially symmetric monopole (or spherically symmetric if $\ell=1$) and an axially symmetric monopole of charge $N-\ell$. This pair of monopoles move in opposite directions along the line passing through the centre of the initial $N$-gon and perpendicular to it. If $\ell=N/2$ then the cyclic symmetry is again extended to dihedral symmetry $D_N.$

An ansatz for solutions of Nahm's equation, for charge $N$ monopoles with $C_N$ symmetry, reduces it to the Toda equation for the affine algebra $A_{N-1}^{(1)}$ \cite{Su2}. In fact, all monopoles with $C_N$ symmetry can be obtained from this Toda reduction \cite{Br}.
The Toda variables, $q_0,...,q_{N-1}$ and $p_1,...,p_N$, are introduced by writing
\be
T_1+iT_2=2q_0E_{N,1}+\sum_{j=1}^{N-1}2q_jE_{j,j+1}, \quad\quad
T_3=-{i}\sum_{j=1}^Np_jE_{j,j},
\label{Toda}
\ee
where $E_{ij}$ is the matrix unit with $i$ denoting the row and $j$ denoting the column where the entry 1 resides, with all other entries equal to zero.
The Einstein summation convention does not apply here, so all sums have been written explicitly.
Nahm data is centred if all three matrices have vanishing trace, which is imposed by requiring that the total momentum for the Toda system, $\sum_{j=1}^Np_j$, is set to zero.
The extension of cyclic symmetry to dihedral symmetry $D_N$ is obtained by a restriction of the $A_{N-1}^{(1)}$ Toda variables via a folding procedure to obtain a non-simply laced Lie algebra from a Dynkin diagram automorphism \cite{BDH}. For even $N>2$ this yields the Toda equations for the algebra $C_{N/2}^{(1)}$, and for odd $N>1$ it is the twisted affine algebra $A_{N-1}^{(2)}.$
Explicitly, the restriction for $D_N$ symmetry may be taken to be, $q_j=q_{N-j}$, for $j=1,..,N-1,$ together with $p_j=-p_{N+1-j}$, for $j=1,..,N.$

The spectral curve for a charge $N$ monopole has genus $(N-1)^2$, so generically Nahm data for $N>2$ cannot be written explicitly in terms of elliptic functions. However, there are exceptions for particularly symmetric monopoles, where the spectral curve is the Galois cover of an elliptic curve. The examples of Nahm data with $N>2$ that were recycled in Section 4 fall into this category. Generically, imposing $C_N$ symmetry reduces the genus of the quotient curve to only $N-1$, and Nahm data is not explicitly available for the geodesics $G_N^\ell$ with $N>2.$ An exception is the planar scattering $G_N^0$, where Nahm data can be written in terms of elliptic functions, and the example $G_3^0$ has been studied in detail \cite{BDH2}. The genus 2 quotient curve for the scattering $G_3^1$ has been investigated using sophisticated methods that include its numerical computation \cite{BDE}.

The rational map description of Euclidean monopoles extends to hyperbolic monopoles \cite{At}, therefore the imposition of cyclic $C_N$ symmetry produces analogous one-parameter families of hyperbolic monopoles, that will also be denoted by $G_N^\ell$, although the interpretation in terms of geodesics and monopole scattering no longer holds. Even when the associated Nahm data is not available to be recycled, the Toda reduction (\ref{Toda}) can still be adapted to obtain hyperbolic monopole data, as follows.

The candidate hyperbolic data is given by $M_i=-iU^\dagger T_iU$, for some $U\in U(N)$, where $T_i$ are taken to have the Toda form (\ref{Toda}), with the Toda variables now constants, corresponding to their evaluation at $s=0.$ Note that there is no requirement here that the Toda solution with the given initial conditions corresponds to Nahm data. This means that the reality condition on $M_i$, that followed automatically from recycling Nahm data, must be imposed by a suitable choice of the matrix $U$, that provides the unitary change of basis. This implies that there exists a symmetric unitary matrix $B=UU^t$ such that
$BT_i=T_i^tB$. This is a linear equation for the matrix $B$ that implies linear relations between the Toda parameters that appear in $T_i$ for the existence of a solution. After imposing these relations the linear system is solved to find $B$ and its Takagi factorization $B=UU^t$ is performed to obtain $U$, and hence the form of $M_i$. Note that the Takagi factorization of a general symmetric matrix $B$ cannot be solved by radicals to obtain $U$, but if $B$ is also unitary (as in the application here) then its Takagi factorization is always solvable by quadratic radicals and an explicit procedure to obtain $U$ is available \cite{Ik}.

To obtain a new one-parameter family of hyperbolic monopoles for some $G_N^\ell$ requires that $N\ge 4$, because all hyperbolic monopoles with $N=3$ can already be obtained as JNR monopoles. The family $G_4^2$, with charge 4 and $D_4$ symmetry, is not within the JNR class. It includes the cubic charge 4 hyperbolic monopole, which is the only point within this one-parameter family where the hyperbolic data is already known. $G_4^2$ will therefore be used as an example to illustrate the Toda reduction procedure described in this section. The Toda folding to extend the $C_4$ symmetry to $D_4$ is $q_3=q_1, p_4=-p_1, p_3=-p_2$. In this case the required solution $B$ to the linear system is 
\be
B=\left(\begin{array}{cccc}
0 & 1 & 0 & 0 
\\
 1 & 0 & 0 & 0 
\\
 0 & 0 & 0 & 1 
\\
 0 & 0 & 1 & 0 
\end{array}\right),
\ee
which imposes the reality conditions $q_2=q_0$ and $p_2=p_1$.
The real symmetric matrices
\bea
& &M_1=
\left(\begin{array}{cccc}
0 & {q_1}  & 0 & -{q_0}  
\\
 {q_1}  & 0 & -{q_0}  & 0 
\\
 0 & -{q_0}  & 0 & {q_1}  
\\
 -{q_0}  & 0 & {q_1}  & 0 
\end{array}\right),\quad
M_2=
\left(\begin{array}{cccc}
-{q_1}  & 0 & -{q_0}  & 0 
\\
 0 & {q_1}  & 0 & {q_0}  
\\
 -{q_0}  & 0 & -{q_1}  & 0 
\\
 0 & {q_0}  & 0 & {q_1}  
\end{array}\right),\nonumber\\
&&M_3=
\left(\begin{array}{cccc}
-{p_1}  & 0 & 0 & 0 
\\
 0 & -{p_1}  & 0 & 0 
\\
 0 & 0 & {p_1}  & 0 
\\
 0 & 0 & 0 & {p_1}  
\end{array}\right),
\eea
are obtained after performing the Takagi factorization of $B$ and the change of basis from the resulting $U.$
Calculating the associated matrices using (\ref{defSA}) yields
\bea
&&
A_1=2q_0p_1
\left(\begin{array}{cccc}
0 & 0 & -1 & 0 
\\
 0 & 0 & 0 & 1 
\\
 1 & 0 & 0 & 0 
\\
 0 & -1 & 0 & 0 
\end{array}\right),\quad
A_2=2q_0p_1
\left(\begin{array}{cccc}
0 & 0 & 0 & 1 
\\
 0 & 0 & 1 & 0 
\\
 0 & -1 & 0 & 0 
\\
 -1 & 0 & 0 & 0 
\end{array}\right),\nonumber\\
&&
A_3=2(q_1^2-q_0^2)
\left(\begin{array}{cccc}
0 & 1 & 0 & 0 
\\
 -1 & 0 & 0 & 0 
\\
 0 & 0 & 0 & 1 
\\
 0 & 0 & -1 & 0 
\end{array}\right),
\eea
with $S=2q_1^2+2q_0^2+p_1^2.$

Setting $p_1=(q_1^2-q_0^2)/q_0$, 
the matrices $r_i=A_i/(2q_1^2-2q_0^2)$ are a basis for the representation $\underline{2}\oplus\underline{2}$, with $S=(q_1^4+3q_0^4)/q_0^2.$
Comparing this normalization of $A_i$ with (\ref{ftor}) requires that $8(q_1^2-q_0^2)=1-\alpha^2$, and equating $S$ to the expression (\ref{f4}) with $d=2$ requires that $3+\alpha^2=4(q_1^4+3q_0^4)/q_0^2.$ The required solution is
\be
\alpha^2=(1+4q_0)^2,\quad q_1=\sqrt{-q_0(1+q_0)},
\ee
for $q_0\in (-\frac{1}{2},0),$ which means that the expression for $p_1$ simplifies to $p_1=-1-2q_0.$

The spectral curve obtained using (\ref{sc}) with this hyperbolic data is
\be
   q_0^{3}\left(\eta^{4}+1\right)\left(\zeta^{4}+1\right) 
   +q_0^{2}\left(\eta^{2}+\zeta^{2}\right) \left(\eta +\zeta \right)^{2}
   +q_0 \eta  \zeta \left(\eta^{2}+3 \eta  \zeta +\zeta^{2}\right) +\eta^{2} \zeta^{2}=0.
   \ee
   In the limit $q_0\to-\frac{1}{2}$ the curve becomes
   $(\eta^4-1)(\zeta^4-1)=0$, which corresponds to 4 monopoles on the vertices of a square at the boundary
   of hyperbolic space
   $|X|=1$, with the monopole positions $(X_1,X_2,X_3)$ given by $(\pm 1,0,0)$ and $(0,\pm 1,0)$. When $q_0=-\frac{1}{4}$ the curve becomes
   \be
   3\left(\eta -\zeta \right)^{4}-{\eta^{4} \zeta^{4}}+6 \eta^{2} \zeta^{2}+{4 \eta  \zeta \left(\eta^{2}+\zeta^{2}\right)}-{1}=0,
   \ee
   which is the curve of the cubic monopole.
   In the limit $q_0\to 0$ the curve becomes  $\eta^2\zeta^2=0$, which is the curve of a pair of axial charge 2 monopoles at the boundary, with positions $(0,0,\pm 1)$.
   
\begin{figure}[!ht]\begin{center}
    \includegraphics[width=1.0\columnwidth]{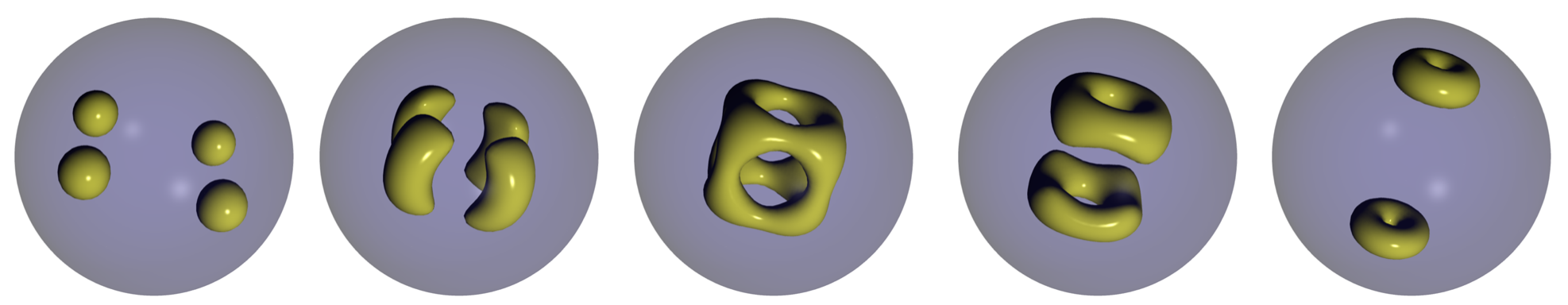}
    \caption{
      Energy density isosurfaces for the family $G_4^2$, with the parameter value from left to right given by $q_0=-0.45,-0.30,-0.25,-0.20,-0.05.$
    }
    \label{fig:sig42}\end{center}\end{figure}

$L$ is obtained from (\ref{defL}) and (\ref{complexeig}) to be   
   \be
   L=\sqrt{-q_0(1+2q_0)}(
   {j}+1,
   {k}+{i},
   {k}-{i},
   {j}-1),
   \ee
allowing the computation of $|\Phi|^2$ using (\ref{Higgs}). Applying the 
Laplace-Beltrami operator to $|\Phi|^2$ produces the energy density isosurfaces
displayed in Figure \ref{fig:sig42}, for a set of increasing values of the parameter $q_0$. This provides a visual representation to complement the spectral curve description. These images clearly match the rational map expectations for the asymptotic configurations at each end of the $G_4^2$ family.

The family $G_N^1$ is of JNR type, obtained by placing one point on the sphere at the North pole and $N$ points with canonical weights at the same latitude on the vertices of a regular $N$-gon. This latitude is the parameter of the family. The example $G_3^1$ has been studied in detail \cite{BCS}. Despite the fact that the family $G_4^1$ can be obtained from JNR data, it seems worthwhile to briefly present the results within the hyperbolic data formulation.

\begin{figure}[!ht]\begin{center}
    \includegraphics[width=1.0\columnwidth]{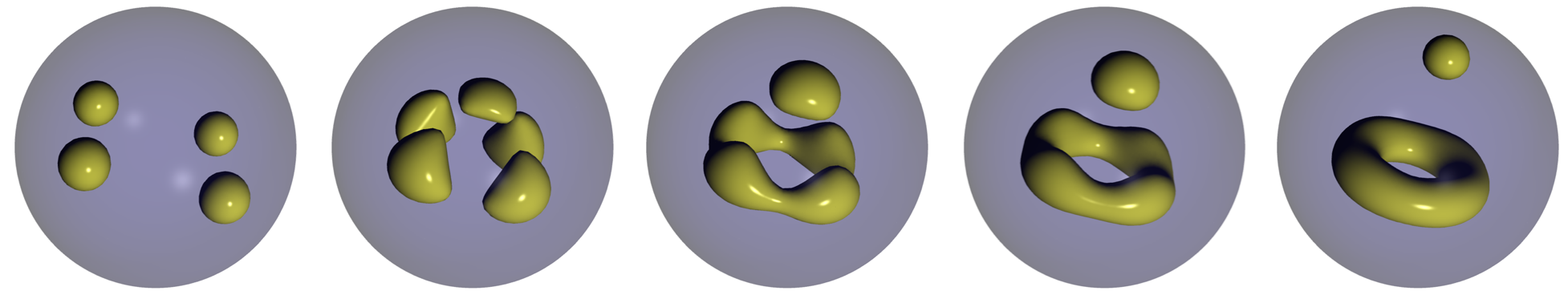}
    \caption{
      Energy density isosurfaces for the family $G_4^1$, with the parameter value from left to right given by $p_1=0.05,0.20,0.25,0.28,0.32.$
    }
    \label{fig:sig41}\end{center}\end{figure}

The required solution $B$ of the linear system for $G_4^1$ is 
   \be
   B=\left(\begin{array}{cccc}
1 & 0 & 0 & 0 
\\
 0 & 0 & 0 & 1 
\\
 0 & 0 & 1 & 0 
\\
 0 & 1 & 0 & 0 
   \end{array}\right),
   \ee
which imposes the reality conditions $q_3=q_2, q_1=q_0, p_4=p_2,$ and therefore $p_3=-p_1-2p_2.$ The Takagi factorization and subsequent change of basis produces
\bea
M_1&=&\sqrt{2}   
\left(\begin{array}{cccc}
0 & 0 & 0 & -q_{0} 
\\
 0 & 0 & 0 & 0 
\\
 0 & 0 & 0 & q_{2} 
\\
 -q_{0} & 0 & q_{2} & 0 
\end{array}\right),\quad
M_2=\sqrt{2}
\left(\begin{array}{cccc}
0 & -q_{0} & 0 & 0 
\\
 -q_{0} & 0 & -q_{2} & 0 
\\
 0 & -q_{2} & 0 & 0 
\\
 0 & 0 & 0 & 0 
\end{array}\right),\nonumber\\
M_3&=&\left(\begin{array}{cccc}
-p_{1} & 0 & 0 & 0 
\\
 0 & -p_{2} & 0 & 0 
\\
 0 & 0 & p_{1}+2 p_{2} & 0 
\\
 0 & 0 & 0 & -p_{2} 
\end{array}\right).
\eea
In this case the associated representation is $\underline{3}\oplus\underline{1}$ with the
solution given by $q_0=\frac{1}{2}\sqrt{1-p_1^2}$, $q_2=\frac{1}{2}\sqrt{(1-p_1)(1-3p_1)}$, $p_2=p_1$, $\alpha^2=(1-4p_1)^2$, for $p_1\in(0,\frac{1}{3})$.
The spectral curve is
\bea
& &(3p_1^4-1)(\eta^4-1)(\zeta^4-1)-2p_1(\eta^2\zeta^2+1)^2(2p_1^2+p_1-2)\\
& &+8p_1(\eta^2+\zeta^2)(p_1^2(\eta^2-\eta\zeta-\zeta^2)-\eta\zeta)
-2p_1^2(3\eta^2-2\eta\zeta-3\zeta^2)(\eta^2-2\eta\zeta-\zeta^2)=0\nonumber.
\eea
In the limit $p_1\to 0$ the curve becomes $(\eta^4-1)(\zeta^4-1)=0$, which corresponds to 4 monopoles on the vertices of a square at the boundary $|X|=1$.
In the limit $p_1\to \frac{1}{3}$ the curve becomes
$\eta(\eta-2\zeta)(\eta^2+4\zeta^2).$ The $\eta$ factor describes a single monopole at $X=(0,0,1)$ and the remaining factor is the spectral curve of a charge 3 axially symmetric monopole at $X=(0,0,-\frac{1}{3}).$
Note that the axially symmetric monopole does not reach the boundary $|X|=1.$ The same phenomenon appears in the $G_3^1$ family of JNR monopoles, where a suggested physical explanation is that the abelian magnetic field on $|X|=1$ can have a vanishing dipole even when a single monopole is at the boundary of $\mathbb{H}^3$ and an axially symmetric monopole is in the interior \cite{BCS}. 
To produce energy density isosurfaces requires calculating $L$ from (\ref{defL}) and (\ref{complexeig}), with the result
\be
L=\sqrt{p_1}(
  0,
  -\sqrt{1-p_1}({i}+{k}),
  \sqrt{2(1-3p_1)}({j}-1),
  -\sqrt{1-p_1}(1+{j})).
  \ee
  Applying the Laplace-Beltrami operator to $|\Phi|^2$, computed using (\ref{Higgs}), yields the isosurfaces displayed in
  Figure \ref{fig:sig41}, for a set of increasing values of $p_1$.
   
The planar family $G_N^0$ is also of JNR type, and is obtained by an appropriate placement of $N+1$ points on the equator of the sphere, with a suitable choice of weights to impose the $D_N$ symmetry. The example $G_3^0$ can be found in \cite{BCS}, and the generalization to any $N$ follows a similar procedure. For completeness, the hyperbolic data for $G_4^0$ is presented below.  

The family $G_4^0$ shares the same dihedral $D_4$ symmetry as the family $G_4^2$, hence the same Toda folding is applied, $q_3=q_1, p_4=-p_1, p_3=-p_2$. The difference between these two families lies in the solution $B$ to the linear system that imposes the reality condition. For $G_4^0$ the solution is
 \be
B= \left(\begin{array}{cccc}
0 & 0 & 0 & 1 
\\
 0 & 0 & 1 & 0 
\\
 0 & 1 & 0 & 0 
\\
 1 & 0 & 0 & 0 
\end{array}\right),
\ee
 which imposes the reality conditions $p_1=0, p_2=0$, and hence $M_3=0$. The Takagi factorization of $B$ and the resulting change of basis gives the other two matrices 
\be
M_1=
\left(\begin{array}{cccc}
-{q_0}  & 0 & 0 & -{q_1}  
\\
 0 & {q_0}  & -{q_1}  & 0 
\\
 0 & -{q_1}  & {q_2}  & 0 
\\
 -{q_1}  & 0 & 0 & -{q_2}  
\end{array}\right),\quad
M_2=
\left(\begin{array}{cccc}
0 & -{q_0}  & -{q_1}  & 0 
\\
 -{q_0}  & 0 & 0 & {q_1}  
\\
 -{q_1}  & 0 & 0 & {q_2}  
\\
 0 & {q_1}  & {q_2}  & 0 
\end{array}\right).
\ee
The solution is $q_1=-q_2=\frac{1}{2},$ with $\alpha^2=4q_0^2$ for $q_0\in (-\frac{1}{2},\frac{1}{2}),$ associated with the representation $\underline{2}\oplus\underline{1}\oplus\underline{1}$ and
the spectral curve 
\be
4q_0^2 \eta  \zeta \left(\eta^{2}-\eta  \zeta +\zeta^{2}\right)
+2q_0 (\eta^{4} \zeta^{4}+1)
+\eta^{4}-\eta^{3} \zeta +\eta^{2} \zeta^{2}-\eta  \,\zeta^{3}+\zeta^{4}
=0.
\ee
 In the limit $q_0\to \pm\frac{1}{2}$ the curve becomes
   $(\eta^4\pm1)(\zeta^4\pm1)=0$, which corresponds to 4 monopoles on the vertices of a square at the boundary circle $X_1^2+X_2^2=1$, with one square rotated by $\pi/4$ compared to the other. With the identification $a=2q_0\in(-1,1)$, this is an obvious generalization of the degenerate charge 2 hyperbolic data presented at the end of Section 4, associated with the representation $\underline{2}$.
In fact, for any $N\ge 2$ the family $G_N^0$ has a similar description, 
associated with the representation $\underline{2}\oplus\underline{1}\oplus ...\oplus\underline{1}$. The solution is degenerate as $M_3=0$, so $A_1=A_2=0$. The only non-zero entries of $A_3$ are in the top left $2\times 2$ block, which is given by
$\frac{1}{2}(1-a^2)r_3$, in terms of the basis (\ref{basis2}) for $\underline{2}.$ The first two entries of the diagonal matrix $S$ are equal to $\frac{1}{2}(1+a^2)$ and the remaining $N-2$ entries are all equal to 1. The associated matrices $M_1$ and $M_2$ are equivalent to the matrices extracted from the constrained ADHM data for the family $G_N^0$ \cite{Co}.

In summary, the three families $G_3^2,G_3^1,G_3^0,$ correspond to the different representations $\underline{2}\oplus\underline{2},\, \underline{3}\oplus\underline{1},\,\underline{2}\oplus\underline{1}\oplus\underline{1},$ and are distinguished within the Toda reduction by different matrices $B$ that impose the reality condition.

\section{Disposable Nahm data}\quad
It appears that many of the currently known examples of explicit Nahm data can be recycled to produce hyperbolic monopole data. However, this is not a general property of Nahm data, as demonstrated in this brief section by presenting an example of Nahm data that is disposable, in the sense that it cannot be recycled to produce hyperbolic data. 
The Nahm data of interest in this section is for the charge $N$ monopole with axial symmetry. For $N=2$ and $N=3$ it has already been shown that this Nahm data can be recycled, because these two cases appeared in the last two examples of Section 4, for particular parameter values within a more general family. It may therefore be surprising that the short calculation below reveals that this Nahm data is disposable for $N\ge 4.$

Take $r_i$ to be a basis for the $N$-dimensional irreducible representation of $\mathfrak{su}(2)$, with $r_1$ and $r_2$ symmetric matrices and $r_3$ an antisymmetric matrix, generalizing the $N=2$ basis given explicitly in (\ref{basis2}). The Nahm data for the axially symmetric charge $N$ monopole is given by \cite{ES}
\be
T_1=-\frac{\pi}{4}\sec(s\pi/2)\,r_1, \qquad
T_2=-\frac{\pi}{4}\sec(s\pi/2)\,r_2, \qquad
T_3=\frac{\pi}{4}\tan(s\pi/2)\,r_3.
\ee
Attempting to recycle produces matrices of the form $M_1=i\beta r_1, \ M_2= i\beta r_2, \ M_3=0,$ which gives
$A_1=0,\, A_2=0,\, A_3=-2\beta^2 r_3,$ and $S=-\beta^2(1-N^2-r_3^2).$ Substituting these expressions into the last equation in (\ref{quadi}), with $\alpha=0$, yields the condition
\be
(1-2\beta^2(N^2-1))r_3=2\beta^2 r_3^3,
\label{axialbeta}
\ee
which requires that $r_3^3$ is proportional to $r_3$. This is true if $N=2$ or $N=3$, since for these two values $r_3^3=-(N-1)^2r_3$, but it is not true for $N>3$. This shows that the axial Nahm data with $N\ge 4$ is disposable. To complete the analysis of the case $N\in\{2,3\}$, the solution of (\ref{axialbeta}) is $\beta=1/\sqrt{4N-4}$, and the first equation in (\ref{quadi}), with $\alpha=0$, reduces to
\be
(N-3)((N-1)^2+r_3^2)=0.
\ee
This is obviously satisfied if $N=3$, and if $N=2$ then $r_3^2=-1$, so it is again satisfied.

The symmetries imposed to make Nahm data tractable often results in only a small number of invariant triplets of matrices from which the Nahm data can be built.
It seems likely that the limited number of invariant triplets of matrices is the property that also allows the Nahm data to be recycled to obtain hyperbolic data. The number of $SO(2)$ invariant triplets of matrices increases with $N$, so the result that the axial Nahm data is recyclable for $2\le N \le 3$, but is disposable for $N\ge 4$, is consistent with this explanation. The expectation is therefore that generic Nahm data is disposable, despite the fact that many of the known explicit examples can be recycled.

\section{Conclusion}\quad
Constrained ADHM data for hyperbolic monopoles has been reformulated, to highlight a resemblance to Nahm data. This new formulation of hyperbolic monopole data, in terms of a triplet of real matrices $M_i$ that satisfy a quartic equation, has revealed a connection with representations of $\mathfrak{su}(2)$. The matrices $M_i$ may be viewed as a prebasis for the basis $r_i$ of the representation, and it might be interesting to investigate the general construction of such a prebasis directly from the representation.

A relationship between hyperbolic data and Nahm data has been investigated, with Toda reduction methods for cyclic Nahm data being adapted to the hyperbolic setting to produce new solutions. A procedure exists \cite{HMM,HS1} to obtain triplets of matrices invariant under a finite subgroup of the rotation group $SO(3)$, in terms of invariant homogeneous polynomials over $\mathbb{CP}^1$. Furthermore, code is publicly available to automatically calculate these matrices given the invariant polynomials as the input data \cite{DH}. These methods were developed for the construction of symmetric Nahm data, but as with the Toda reduction, they may be adapted to construct symmetric hyperbolic monopole data.

Conceptually, the underlying reason why a variety of explicit $SU(2)$ charge $N$ hyperbolic monopole data can be obtained is that the curvature of hyperbolic space is tuned so that these hyperbolic monopoles correspond to $SU(2)$ Yang-Mills instantons with instanton number $N.$ This allows the ADHM construction of instantons to be adapted to hyperbolic monopoles in the simplest fashion.
There are a discrete set of curvatures for hyperbolic space where charge $N$ hyperbolic monopoles correspond to instantons with charge $mN$, for $m$ a positive integer \cite{At}. As the instanton number increases with $m$ it becomes more difficult to obtain explicit data, but it would be interesting to extend the $m=1$ results in this paper to $m>1.$ Another avenue for future research is to consider this approach for gauge groups beyond $SU(2).$


\begin{thebibliography}{99}

\bibitem{At}
M.F. Atiyah, {Magnetic monopoles in hyperbolic spaces},
in \textit{M. Atiyah: Collected Works, vol. 5},
Oxford, Clarendon Press, 1988.  

\bibitem{Nahm} W. Nahm, {The construction of all self-dual
multimonopoles by the ADHM method}, in \textit{Monopoles in Quantum Field
Theory}, eds. N.S. Craigie, P. Goddard and W. Nahm, Singapore,
World Scientific, 1982.   


\bibitem{ADHM} M.F. Atiyah, N.J. Hitchin, V.G. Drinfeld and Yu.I. Manin,
Construction of instantons,
\textit{Phys. Lett.} \textbf{A65}, 185 (1978).

\bibitem{MS}
N.S. Manton and P.M. Sutcliffe,
Platonic hyperbolic monopoles,
\textit{Commun. Math. Phys.} \textbf{325}, 821 (2014).

\bibitem{Hi1} N.J. Hitchin, 
Monopoles and geodesics,
\textit{Commun. Math. Phys.} \textbf{83}, 579 (1982).

\bibitem{JNR} R. Jackiw, C. Nohl and C. Rebbi, 
Conformal properties of pseudoparticle configurations,
\textit{Phys. Rev.} \textbf{D15}, 1642 (1977).

\bibitem{CF} E. Corrigan and D.B. Fairlie,
Scalar field theory and exact solutions to a classical $SU(2)$
gauge theory,
\textit{Phys. Lett.} \textbf{B67}, 69 (1977).

\bibitem{BCS}
  S. Bolognesi, A. Cockburn and P.M. Sutcliffe,
Hyperbolic monopoles, JNR data and spectral curves,
\textit{Nonlinearity} {\bf 28}, 211 (2015).

\bibitem{Su1} P.M. Sutcliffe,
  Spectral curves of hyperbolic monopoles from ADHM,
  \textit{J. Phys. A: Math. Theor.} \textbf{54}, 165401 (2021).

\bibitem{Lee} H.C. Lee,
Eigenvalues and canonical forms of matrices with quaternion coefficients,
\textit{Proc. Roy. Irish Acad.} \textbf{A52}, 253 (1948).

\bibitem{HMM} N.J. Hitchin, N.S. Manton and M.K. Murray,
Symmetric monopoles,
 \textit{Nonlinearity} \textbf{8}, 661 (1995).

\bibitem{HS2} C.J. Houghton and P.M. Sutcliffe, 
Octahedral and dodecahedral monopoles,
\textit{Nonlinearity} \textbf{9}, 385 (1996).

\bibitem{HS1} C.J. Houghton and P.M. Sutcliffe,
Tetrahedral and cubic monopoles,
 \textit{Commun. Math. Phys.} \textbf{180}, 343 (1996).

\bibitem{HS3} C.J. Houghton and P.M. Sutcliffe,
Monopole scattering with a twist,
\textit{Nucl. Phys.} \textbf{B464}, 59 (1996).

\bibitem{BDH2}
  H.W. Braden and L. Disney-Hogg,
 Towards a classification of charge-3 monopoles with symmetry,
 \textit{Lett. Math. Phys. } \textbf{113}, 87 (2023).
 
\bibitem{BPP} S.A. Brown, H. Panagopoulos and M.K. Prasad,
Two separated $SU(2)$ Yang-Mills-Higgs monopoles in the
Atiyah-Drinfeld-Hitchin-Manin-Nahm construction,
\textit{Phys. Rev.} \textbf{D26}, 854 (1982).

\bibitem{Do} S.K. Donaldson, 
Nahm's equations and the classification of monopoles,
\textit{Commun. Math. Phys.} \textbf{96}, 387 (1984).

\bibitem{Ma1} N.S. Manton, 
A remark on the scattering of BPS monopoles,
\textit{Phys. Lett.} \textbf{B110}, 54 (1982).

\bibitem{Su2} P.M. Sutcliffe,
 Seiberg-Witten theory, monopole spectral curves and affine Toda solitons,
\textit{Phys. Lett.} \textbf{B381}, 129 (1996).

\bibitem{Br} H.W. Braden,
  Cyclic monopoles, affine Toda and spectral curves,
  \textit{Commun. Math. Phys.} \textbf{308}, 303 (2011).

\bibitem{BDH}
  H.W. Braden and L. Disney-Hogg,
  Dihedrally symmetric monopoles and affine Toda equations,
  \textit{J. Phys. A: Math. Theor.} \textbf{57}, 495207 (2024).

\bibitem{BDE}
  H.W. Braden, A. D'Avanzo and V.Z. Enolski,
  On charge-3 cyclic monopoles,
  \textit{Nonlinearity} \textbf{24}, 643 (2011).

 \bibitem{Ik} Kh.D. Ikramov,
  Takagi’s decomposition of a symmetric unitary matrix as a finite algorithm,
    \textit{Comput. Math. and Math. Phys.} \textbf{52}, 1 (2012). 
  
\bibitem{Co}  A. Cockburn,
  Symmetric hyperbolic monopoles,
  \textit{J. Phys. A: Math. Theor.} \textbf{47}, 395401 (2014).

\bibitem{ES} N. Ercolani and A. Sinha,
  Monopoles and Baker functions,
  \textit{Commun. Math. Phys.} \textbf{125}, 385 (1989).

\bibitem{DH} L. Disney-Hogg, Symmetries of Riemann surfaces and magnetic monopoles, Ph.D. thesis, University of Edinburgh (2023).
\url{https://github.com/DisneyHogg/Riemann_Surfaces_and_Monopoles}

\end{thebibliography}
\end{document}